\documentclass[twocolumn]{aastex7}

\newcommand{\pc}{\>{\rm pc}}
\newcommand{\kpc}{\mbox{$\>{\rm kpc}$}} 

\newcommand{\Gyr}{\mbox{$\>{\rm Gyr}$}}

\newcommand\degrees{^\circ}
\newcommand{\avg}[1]{\mbox{$\left<{#1}\right>$}}

\newcommand{\feh}{\mbox{$\rm [Fe/H]$}}

\newcommand\gaia{{\it Gaia}}

\begin{document}

\title{Spiral Structure Diversity in Milky Way Analogs from TNG50: The Role of Gas and Disk Dynamics}

\author[orcid=0000-0002-6549-7455,sname='Ghosh']{Soumavo Ghosh}
\affiliation{Department of Astronomy, Astrophysics and Space Engineering, Indian Institute of Technology Indore, India - 453552}
\email[show]{soumavo@iiti.ac.in}  

\author[orcid=0000-0003-2676-8344,gname=Bosque, sname='D'Onghia']{Elena D'Onghia} 
\affiliation{Department of Physics, University of Wisconsin- Madison, Madison, WI 53706, USA}
\affiliation{Department of Astronomy, University of Wisconsin- Madison, Madison, WI 53706, USA}
\email[show]{edonghia@astro.wisc.edu}

\begin{abstract}
The generation of spiral arms and the mechanisms controlling their properties within a realistic cosmological framework - the complete understanding is still beyond our grasp. 
Using a statistically significant sample of Milky Way– and Andromeda–like (MW/M31) analogs from the high-resolution TNG50 cosmological simulation, we carry out the first systematic investigation of spiral-arm formation, their observable properties, and the underlying physical drivers. The selected analogs predominantly exhibit two-armed ($m = 2$) spirals in both stars and gas, while the gaseous disks often display stronger, more tightly wound, and multi-armed patterns ($m>2$). Spiral features appear across stellar populations of different ages, confirming their density-wave nature and producing coherent spirals in both metallicity and mean stellar age distributions—consistent with recent Gaia observations of the Milky Way. 
Our analysis reveals a diverse dynamical scenario for spiral generation: gas content, disk coldness, and shear jointly regulate the growth and morphology of spiral perturbations. We find that the gas content and the dynamical coldness of the disk jointly regulate spiral growth: galaxies with higher gas fractions and colder disks develop more prominent spirals. The measured relation between spiral pitch angle and disk shear shows significant scatter around the analytic prediction, likely due to the combined influence of bars, gas inflows, and feedback. These results demonstrate that spiral density waves can persist in fully cosmological disks, linking internal dynamical processes to galaxy assembly and offering testable predictions for present and future surveys such as JWST and Roman.

\end{abstract}


\section{Introduction}
\label{sec:Intro}
Spiral arms are among the most common non-axisymmetric features in disk galaxies—second only to stellar bars—in both the local Universe  
\citep[e.g see][]{Elmegreenetal2011,Yuetal2018,Savchenkoetal2020} and in  galaxies observed up to redshift $Z_r \sim$ 1.8 \citep[e.g.][]{Elmegreenetal2014,Willetetal2017,Hodgeetal2019}. However, despite decades of theoretical and numerical work, the physical origin and persistence of spiral structure in galaxies that evolve within a full cosmological context, subject to accretion, feedback, and mergers, remain poorly quantified.
Spirals often co-exist with other non-axisymmetric structures, such as an $m=2$ stellar bar and/or an $m=1$ lopsidedness \citep[e.g., see][]{RixandZaritsky1995, Bournaudetal2005, Butaetal2010,Zaritskyetal2013,Kruketal2018}. The Milky Way itself hosts a stellar bar and multiple spiral arms \citep[e.g. see][]{Weinberg1992, Gerhard2002}. A wide variety of physical mechanisms have been proposed to explain the origin of spiral arms, including bar-driven spirals \citep[e.g.,][]{Salo2010}, swing amplification of perturbations \citep{GoldreichLyden65, JulainToomre66, Toomre81, Donghia2013}, tidal interactions \citep[e.g.,][]{Toomre1972, Dobbsetal2010}, mutual interactions with other spirals \citep{Masset1997}, 
recurrent groove modes \citep{SellwoodLin1989, Sellwood2012, SellwoodandCarlberg2019} and manifold-driven spirals \citep{Athanassoula2012}. 
The physical nature of spiral arms--whether they are long-lived patterns or transient features--has been a subject of long-standing debate.The quasi-stationary density wave theory \citep[e.g.][]{LinandShu1964,LinandShu1966,Shu2016} posits that spirals rotate with a constant pattern speed distinct from the disk, a view supported and challenged by various observational studies \citep[e.g. see][]{Schinnereretal2017,Chandaretal2016,Shabanietal2018}.  
Alternatively, other studies argued that spiral arms as transient features, arising by the combined effects of epicyclic motion of stars, shear due to differential rotation, and amplified by disk self-gravity, a process known as swing amplification \citep[e.g.][]{GoldreichLyden65, JulainToomre66, Toomre81}
\par
 Irrespective of their origin or longevity, spiral arms significantly influence the structure and dynamics of disk galaxies. They facilitate angular momentum redistribution \citep{LyndenBellKalnajs1972}, drive radial migration of stars without disk heating \citep[e.g. see][]{SellwoodBinney2002, Roskaretal2008, SchonrichBinney2009, VeraCiro14, VeraCiro16, VeraCiro16b}, induce vertical breathing motions \citep[e.g.][]{Faureetal2014, Debattista2014,Donghia16, Ghoshetal2022,Kumaretal2022,Khachaturyantsetal2022}, and even contribute to outer stellar disk heating \citep[e.g.][]{Sahaetal2010} and driving a large-scale gradient in the bulk radial velocity \citep[e.g.][]{Siebertetal2011,Siebertetal2012}. These dynamical processes are strongly modulated by the properties of spiral arms, such as strength, radial extent, multiplicity, and pattern speed, all of which are in turn shaped by the underlying mechanism of formation of the arms and their lifetime. A detailed understanding of the mechanisms that govern the formation, persistence, and characteristics of spiral arms is therefore essential for building a coherent picture of secular evolution of disk galaxies.
\par
Previous efforts have addressed the dynamical role of various mechanisms in shaping spiral structure, using both theoretical models and observational diagnostics. For example, \citet{Donghia2013} showed that the spiral structures can persist (at least, statistically) over long timescales as a result of the \textit{non-linear} response of stellar disks to perturbations by massive clumps such as giant molecular clouds. In addition, \citet{SahaandElmegreen2016} showed that bulges play a pivotal role in the sustenance of spirals in $N$-body simulations. The interstellar gas which is a dynamically-cold component, plays a key role on spiral generation and survival by cooling the stellar disk and facilitating the generation of fresh spiral waves \citep{SellwoodCarlberg1984}, making the swing-amplified stellar arms broader \citep{Jog1992} in agreement with observations, and helping the spiral density waves to survive for a longer time \citep[e.g.][]{GhoshJog2015, GhoshJog2016}. Past semi-analytical works further showed that the disk thickness shows an opposite dynamical effect (as compared to gas) in generation of spirals in disk galaxies \citep[e.g. see][]{GhoshandJog2018,GhoshandJog2022}. The spiral arm multiplicity is shown to depend on the disk-mass fraction \citep[wrt. to the dark matter halo mass, e.g. see][]{D'Onghia2015,Fujiietal2018}. The dependence of the spiral pitch angle on the differential rotation (or equivalently, the disk shear) have been investigated from isolated $N$-body simulations \citep[e.g. see][]{Grandetal2013,Michikoshi2016,Fujiietal2018} as well as from observations \citep[e.g. see][]{Seigar2005,Seigaretal2006}. 
While these studies, taken together, provide valuable insights into the mechanisms of spiral generation and longevity, they often rely on idealized, isolated galaxy models or lack the self-consistent treatment of stars, gas, and cosmological context. What remains missing is a systematic investigation of spiral arm generation and its associated properties in galaxies where stellar and gaseous disks evolve naturally within a fully cosmological framework, subject to realistic accretion histories, mergers, and feedback. In this work, we aim to fill this gap by analyzing the diversity of spiral structures in a statistically meaningful sample of Milky Way– and M31–like galaxies (hereafter, MW/M31 analog) drawn from the TNG50 cosmological simulation. This study connects spiral morphology, strength, and multiplicity to the physical conditions of disks shaped by their cosmological histories, bridging local dynamical theory with galaxy evolution, in contrast with previous studies which mainly relied on idealized, isolated disks.
\par
The rest of the paper is organised as follows. Section.~\ref{sec:data_TNG50} provides a brief description of the TNG50-1 cosmological simulations and the details of the selected sample of MW/M31 analogs used in this paper. Section~\ref{sec:spiralProps} contains the results pertaining to quantification of spiral properties in the MW/M31 analogs. Section~\ref{sec_physicalImpacts} provides the details of the underlying physical factors governing the diversity in spiral properties. Section~\ref{sec:discussion} contains the discussion and  Sect.~\ref{sec:conclusion} summarizes the main findings of this work.

\section{TNG50-1 cosmological simulation : MW/M31 analogs}
\label{sec:data_TNG50}

\begin{figure*}[]
\includegraphics[width=1.025\linewidth]{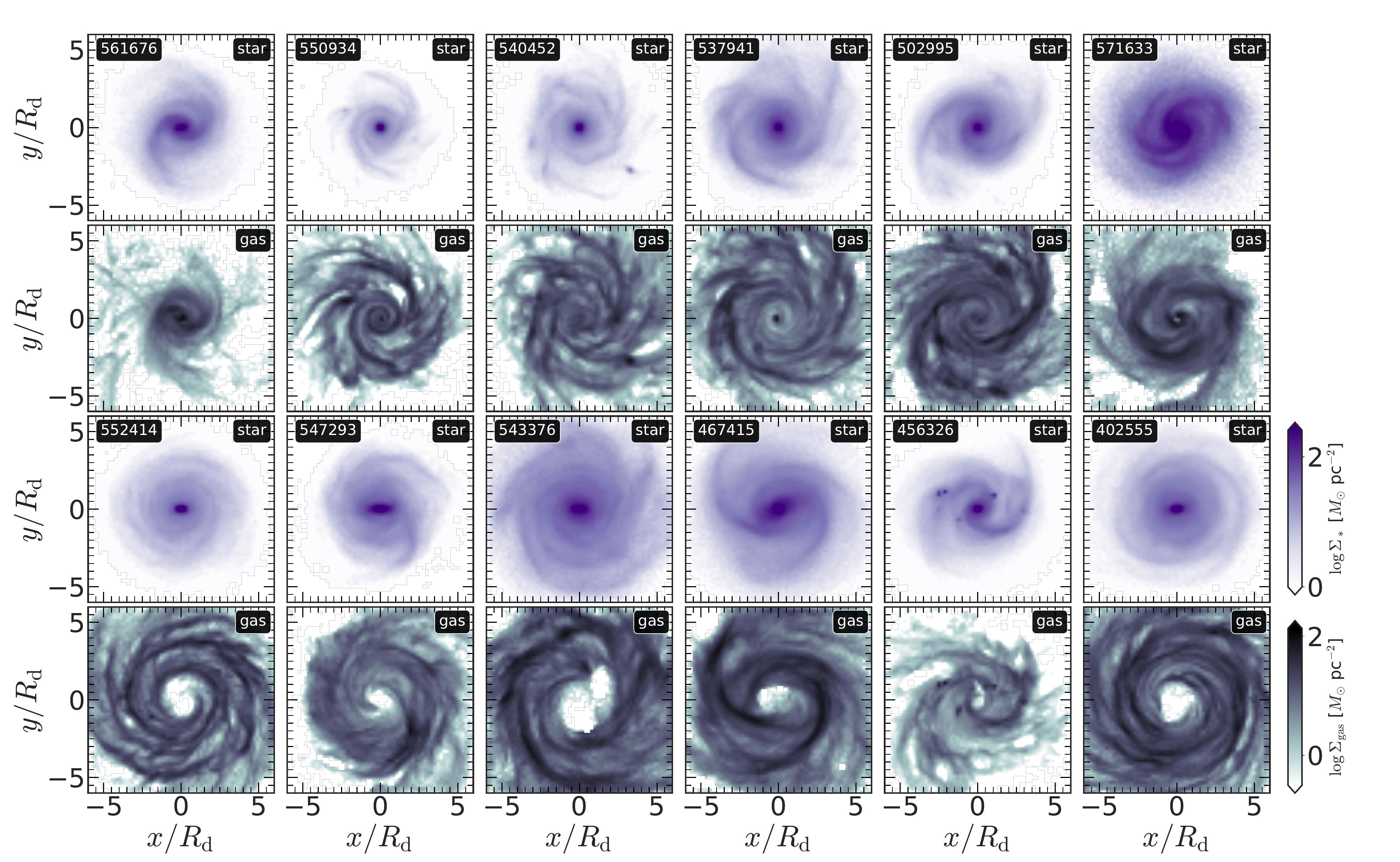}
\caption{ \textit{Diverse spiral morphology in MW/M31 analogs:} Face-on ($x-y$-plane) density distribution of the stars (\textit{first and third rows}) and gas (\textit{second and forth rows}) for a sample of non-barred (\textit{top two rows}) and spirals+bar (\textit{bottom two rows}) galaxies, considered for this work. $R_{\rm d}$ denotes the scale length of the stellar disk, and the values are taken from \citet{Pillepichetal2023}. The color bars show the corresponding surface density (in logarithmic scale). The $\texttt{SubHaloID}$ is mentioned in each of the sub-panels. Both the non-barred and the spirals+bar samples show a wide variety in spiral morphology: from grand-design spirals to multi-armed flocculent spirals. }
\label{fig_densXYmap_combined}
\end{figure*}

Milky Way–mass systems occupy the mass range where disks are marginally stable and most sensitive to feedback, making them natural laboratories for studying spiral dynamics in realistic environments.
In this study, we analyze spiral features in a sample of Milky Way and Andromeda analogs identified within the cosmological simulation TNG50-1 
\footnote{\url{https://www.tng-project.org/data/milkyway+andromeda/}}.These analogs represent central disk galaxies selected to resemble the MW/M31 in terms of stellar mass, morphology, and local environment, as detailed in \citet{Pillepichetal2023}. Below, we briefly describe the simulation parameters and the sample selection criteria used in this work.
\par
TNG50-1 is the highest-resolution run of the IllustrisTNG suite of cosmological magnetohydrodynamical simulations of galaxy formation \citep{Nelsonetal2019a,Nelsonetal2019b, Pillepichetal2019}. TNG50-1 simulates the formation and evolution of galaxies in a 51.7 comoving Mpc$^3$ box from redshift $Z_r$ $\approx$ 127 to $Z_r =0$. It is run with the moving-mesh code AREPO~\citep{Springeletal2010} and uses the fiducial TNG galaxy formation model~\citep{Weinbergeretal2017,Pillepichetal2018} which includes prescriptions for star formation, feedback, and black hole growth, with a mass resolution of $m_{\rm baryon} = 8.5 \times 10^4 M_{\odot}$, $m_{\rm DM} = 4.5 \times 10^5 M_{\odot}$; and a spatial resolution of star-forming gas of $\sim 150 \pc$ \citep{Pillepichetal2023}. We refer to the galaxy at the potential minimum as the central; others are considered satellites. 
  The TNG50 simulations adopt a cosmology consistent with the Planck 2015 analysis \citep{Planck2016} : $\Omega_{\lambda} = 0.6911$, $\Omega_m = 0.3089$, $\Omega_{\rm b} = 0.0486$ and $h = 0.6774$.
\par

The selection criteria for the MW/M31 analogs at $Z_r = 0$ include the following: (i) the stellar mass of the galaxy is in the following range: $\rm M_*(<30 kpc) = 10^{10.5-11.2}M_\odot$; (ii) a disk-like stellar morphology; (iii) no other galaxy with stellar mass $> 10^{10.5} M_\odot$ is located within the $500$~ kpc distance; and (iv) the total mass of the halo host is smaller than that typical of massive groups $< 10^{13}M_\odot$\citep[for further details, see][]{Pillepichetal2023}. The TNG50-1 simulation contains 198 such MW/M31 analogs. We further divided the analogs into two sub-samples: \textit{non-barred} and \textit{spirals+bar} systems. Barred galaxies were identified using published bar classifications from \citet{Zanaetal2022} and \citet{Rosas-Guevaraetal2022}, supplemented by a consistency check based on the $m=2$ Fourier amplitude and its phase constancy within $5-8 \degrees$ in the bar region \citep{GhoshandDiMatteo2024}, as well as the presence of a butterfly pattern in the stellar mean radial velocity field, which is a kinematic signature of bar-induced orbital motions \citep{Ghoshetal2025}.
\par 
Many of the TNG50 MW/M31 analogs host dense clump-like structures that are common in high-resolution cosmological simulations. These clumps introduce strong local density fluctuations, which can contaminate Fourier-based quantification of bar and spiral properties. To address this, we develop the following strategy: we only consider stars within $|z| \leq 5 \kpc$ from the disk mid-plane ($z=0$) for subsequent calculations (including the Fourier decomposition). This helps to mitigate clump effects in many MW/M31 analogs. Systems where the contamination remained unmanageable were excluded from the final sample. This cleaning step ensures that the measured Fourier amplitudes reflect genuine global disk structures rather than transient density fluctuations from local clumps, which are common in high-resolution cosmological runs.
\par 
Our initial, visually-confirmed (and also, clump-free) MW/M31 catalog comprises 32 non-barred galaxies and 65 spiral + bar galaxies. Fig.~\ref{fig_densXYmap_combined} shows the diverse spiral morphology in stars and gas, for a sample of non-barred and spirals+bar galaxies from our initial sample of MW/M31 analogs. 

\section{Quantifying the spiral properties}
\label{sec:spiralProps}

A mere visual inspection of Fig.~\ref{fig_densXYmap_combined} reveals that the selected sample of the MW/M31 analogs display a wide variety in spiral morphology, starting from grand-design spirals to intermediate spirals to  multi-armed flocculent spirals. In order to determine the spiral strength as well as the dominant arm multiplicity, $m$, we computed the radial variations ($0-6 R_{\rm d}$) of different azimuthal Fourier components of the mass distribution in the disk using

\begin{equation}
A_m/A_0 (R)= \left| \frac{\sum_j m_j e^{im\phi_j}}{\sum_j m_j}\right|\,,
\label{eq:fourier_calc}
\end{equation}
 where $A_m$ denotes the coefficient of the $m$th Fourier moment of the density distribution, $m_j$ is the mass of the $j$th particle, and $\phi_j$ is the corresponding azimuthal angle in the cylindrical coordinates \footnote{The summation runs over all the particles within the radial  annulus $[R, R+\Delta R]$, with $\Delta R = 0.5 \kpc$.}. Here, we considered $m=1$ to $m=8$ where $m=1$ represents a one-armed spiral and/or an $m=1$ lopsidedness. We find that, indeed some MW/M31 analogs harbor an $m=1$ lopsided distortion in their stellar density field. As demonstrated in past literature \citep[e.g. see][]{Ghoshetal2022a}, in presence of an $m=1$ lopsided distortion, consideration of a density-weighted center in Fourier decomposition analysis is extremely important. Therefore, we centered the radial annuli (used in the Fourier decomposition, Eq.~\ref{eq:fourier_calc}) wrt. the density-weighted center of the underlying mass distribution. Following \citet{CasertanoandHut1985},  we calculated the density-weighted center of stars using
\begin{equation}
{\bf{x}}_{d,j} = \frac{\sum_{i} {\bf x}_i \rho^{(i)}_j}{\sum_{i} \rho^{(i)}_j}\,,
\end{equation}
\noindent where ${\bf x}_i$ is the three-dimensional position vector for the \textit{i}th particle, and $ \rho^{(i)}_j$ is the density estimator of order \textit{j} around the  \textit{i}th particle, and is evaluated as
\begin{equation}
\rho_j = \frac{j-1}{V(r_j)}m_i\,.
\end{equation}
\noindent Here, $m_i$ is the mass of the particle, $r_j$ is the distance of the \textit{j}th particle from the particle around which the local density is estimated. Here, we chose $j=6$, as prescribed by \citet{CasertanoandHut1985}.This value balances accuracy and noise in estimating local density. We note that for an accurate measurement of Fourier coefficients in different radial annuli, it is crucial to determine the center accurately. Hence, the usage of a density-weighted center is required in the presence of a $m=1$ lopsided distortion, even when computing the Fourier coefficient other than $m=1$.
\par
In this work, we are interested in studying the spiral properties in MW/M31 analogs with prominent spirals. While comparing the radial profiles of $A_2/A_0$, we find that in many cases, the values of $A_2/A_0$ lie below 0.1, and the corresponding spirals are weaker in the stellar density distribution. We define `prominent spirals' where $A_2/A_0 > 0.1$ in the radial range of our interest. Applying this (operational) criterion for prominent spirals, we are left with 16 non-barred and 27 spirals+bar MW/M31 analogs. The subsequent analyses will be done on these samples of MW/M31 analogs harboring prominent spirals. Furthermore, we checked that in all of our selected spirals+bar samples, the $m=2$ bar resides within $1 R_{\rm d}$, and spirals are often prominent in the radial range $R \in [2-4] R_{\rm d}$. Therefore, to avoid any spatial overlap/contamination (due to a bar) in quantifying spirals' properties, we define our radial range of interest as $R \in [2-4] R_{\rm d}$. We verified that varying the amplitude threshold from 0.1 to 0.08 or 0.12 does not alter the statistical trends. The final analysis thus includes 43 MW/M31 analogs (16 non-barred and 27 barred), all hosting well-defined spiral structure.

\subsection{Spiral strength and arm multiplicity}
\label{sec:arm_multiplicity}

\begin{figure}
\includegraphics[width=1.02\linewidth]{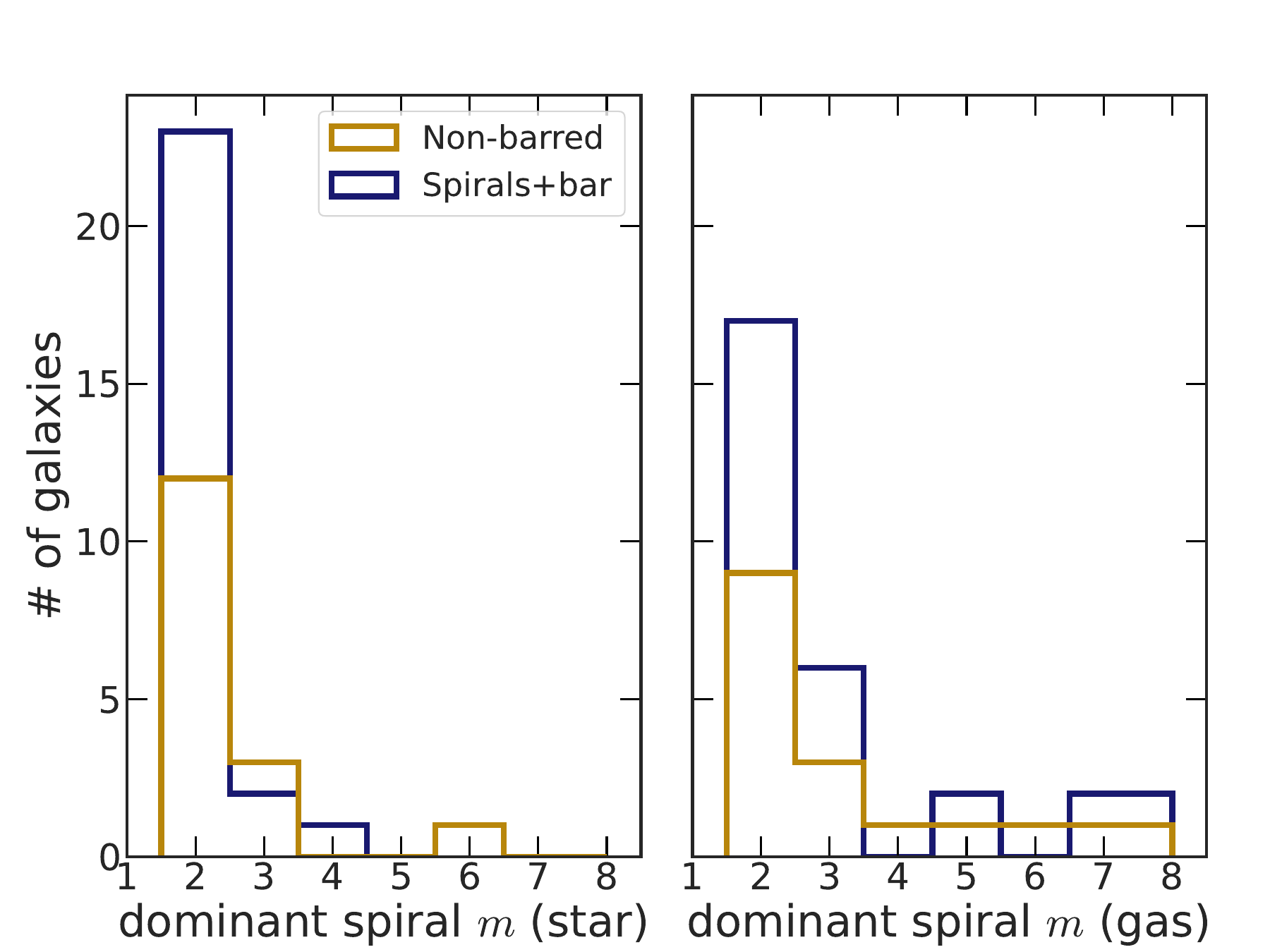}
\caption{Distribution of the most dominant spiral $m$ (arm multiplicity) for both stars (\textit{left panel}) and gas (\textit{right panel}). For further details, see the text. Non-barred galaxies are shown in blue while the spirals+bar galaxies are shown in yellow. }
\label{fig:mostprominent_spiral}
\end{figure}

\begin{figure}
\includegraphics[width=0.95\linewidth]{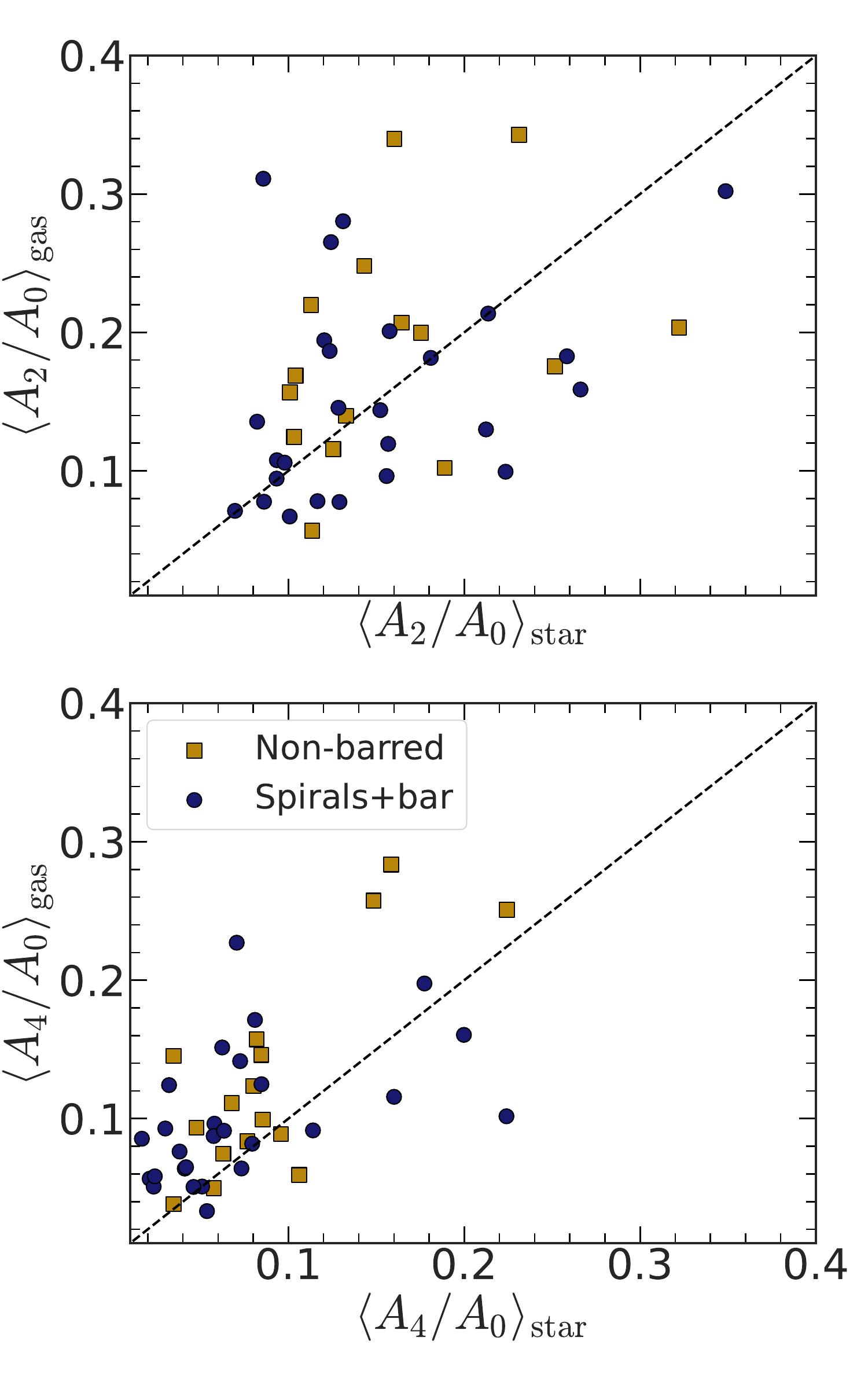}
\caption{Comparison of the spiral $m=2$ (\textit{top panel}) and $m=4$ (\textit{bottom panel}) arm strengths, for the stellar and the gas particles. The black dashed line in each panel denotes the 1:1 correspondence. The blue filled circles denote the spirals+bar galaxies whereas the yellow filled squares denote the non-barred galaxies from our selected MW/M31 analogs. }
\label{fig:armStrength_m2_m4_comparison_starsvsgas}
\end{figure}

In order to quantify the spiral properties (strength, arm multiplicity), we first calculated radial profiles of Fourier coefficients for $m=1$ to $m=8$ using Eq.~\ref{eq:fourier_calc}, for stellar and gas particles separately. We find that in several cases, the corresponding radial profiles of the Fourier amplitudes are noisy; thereby requiring a better technique of quantifying the spiral strength (instead of taking a `single' value of $A_m/A_0$ at a given radial location). To quantify the average spiral strength across a region, we computed $\avg{A_m/A_0}$, the radially averaged Fourier amplitude.  Following \citet{SahaandNaab2013}, we calculated the integrated contribution of different Fourier components over a radial range as

\begin{equation}
\avg{A_m/A_0} = \frac{1}{(R_{\rm max}- R_{\rm min})} \int_{R_{\rm min}} ^{R_{\rm max}} A_m/A_0 (R) \ dR\,,
\end{equation}
\noindent where $R_{\rm min}$ and $R_{\rm max}$ denote the boundary points of the integration. Using  $R_{\rm min} = 2 R_{\rm d}$ and $R_{\rm max} = 4 R_{\rm d}$ (for details, see section~\ref{sec:spiralProps}), we calculated the values of $\avg{A_m/A_0}$ (for $m=1$ to $m=8$), both for stars and gas. This approach ensures robustness against local noise in amplitude profiles. For a given species of particles (stars or gas), we define the dominant spiral $m$ as the value of Fourier moment $m$ for which the value of $\avg{A_m/A_0}$ is maximum. Fig.~\ref{fig:mostprominent_spiral} shows the corresponding distribution of dominant spiral $m$ for stars and gas, computed for our final samples of non-barred and spirals+bar MW/M31 analogs. 
For the stellar disk, a dominant $m=2$ spirals is seen (see left panel of Fig.~\ref{fig:mostprominent_spiral}), for the large fraction of samples (both non-barred and spirals+bar) with occasional dominant multi-armed spirals (i.e. $m > 2$). A similar trend also holds true for the spirals in gas density distribution. However, gas disks show a larger fraction MW-analogs harboring multi-armed spirals as compared to stars (compare left and right panels of Fig.~\ref{fig:mostprominent_spiral}). A plausible explanation is that the gas disk, being dynamically colder than the stellar disk, is more responsive to perturbations. This allows higher-$m$ spirals to form more easily in gas, especially in the outer disk where shear (due to galactic differential rotation) is lower and gas is less stabilized. This may reflect the combined effect of disk stability, external accretion, and feedback-driven structures.
\par
While the spiral arm multiplicity in stars and gas show a variety in their occurrence, it is also worthy investigating how the strengths of spirals in stars and gas compare within the spatial extent considered here. In Fig.~\ref{fig:armStrength_m2_m4_comparison_starsvsgas}, we examined the average $m= 2,4$ Fourier coefficients ($\avg{A_m/A_0}$), calculated separately for stars and gas, within our chosen spatial extent ($R \in [2-4] \ R_{\rm d}$). We find that, in majority of the cases ($\sim 75$ \% for non-barred and $\sim 65$ \% for spirals+bar), the $m=2 $ spiral strength in gas disk is larger (i.e. larger $\avg{A_m/A_0}$ values) when compared to the $m=2$ spiral strength in the stellar disk. 
This trend becomes even stronger for $m=4$ spirals, irrespective of presence of an $m=2$ stellar bar. The physical interpretation is the following. The interstellar gas is dynamically colder when compared to stars. In other words, the sound speed of gas is much lesser when compared to the velocity dispersion of stars \citep[e.g. see][]{SellwoodBinney2002}. Consequently, the gas disk is more susceptible to gravitational instability and would respond more strongly to any perturbation (as compared to stars).
This trend is consistent with theoretical expectations and past simulation results showing that the colder, dynamically responsive gas component typically exhibits stronger and more coherent spiral features than the stellar disk. In linear theory, gas responds more efficiently to perturbations due to its lower sound speed, allowing it to amplify density waves with minimal damping. This has been supported by pure $N$-body simulations that included low-dispersion components \citep[e.g.][]{SellwoodCarlberg1984}, and by semi-analytic work illustrating the gas-driven enhancement of swing amplification \citep[e.g.][]{Jog1992}. More recent studies using $N$-body+SPH or grid-based hydrodynamic simulations confirm that gas enhances and sustains spiral structure by cooling the disk and reinforcing density contrasts \citep[e.g.][]{Dobbsetal2010}. Semi-analytic models further demonstrate that the presence of a cold gas disk can broaden and lengthen the lifetime of spiral arms \citep{GhoshJog2015, GhoshJog2016}, providing additional support for our findings.

\subsection{Spiral pitch angle}
\label{sec:pitch_angle}

\begin{figure}
\includegraphics[width=0.95\linewidth]{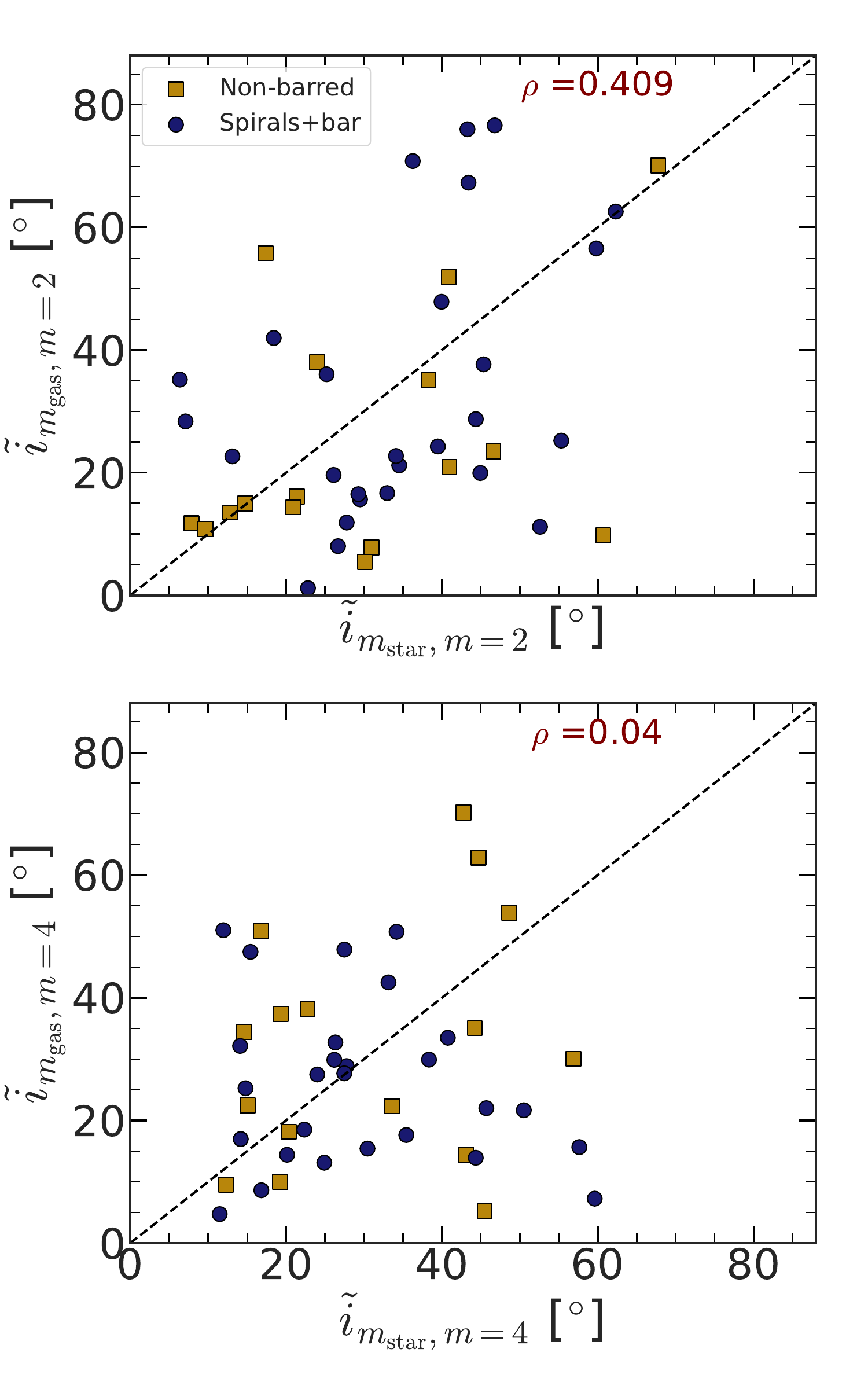}
\caption{Comparison of the average pitch angle ($\avg{i_m}$) of the $m=2$ spirals (\textit{top panel}) and the $m=4$ spirals (\textit{bottom panel}), calculated separately for the stars and gas. The black dashed line in each panel denotes the 1:1 correspondence. The blue filled circles denote the spirals+bar galaxies whereas the yellow filled squares denote the non-barred galaxies from our selected MW/M31 analogs. }
\label{fig:pitchangl_m2_m4_comparison_starsvsgas}
\end{figure}

%
\par%
Another property of interest is the pitch angle of spirals. Following \citet{BinneyTremaine2008}, at radius $R$, we define the pitch angle $i_m (R)$ for a spiral arm with $m$ as 
 \begin{equation}
 \cot{i_m} (R) = \left|R \frac{d \varphi_m(R)}{dR} \right|\,,
 \label{eq:pitch_angle_defn}
 \end{equation}
 \noindent where $\varphi_m$ is the phase angle of the $m$th Fourier moment. In general, the pitch angle varies with the radius. Furthermore, a lower pitch angle indicates a more tightly wound spiral \citep[for further details, see][]{BinneyTremaine2008}. As the spatial resolution of TNG50 ($\sim$150 pc) limits small-scale trends, we report median pitch angles measured in a narrow bin around $R = 2.2\,R_d$, consistent with previous analytic studies.
 \par 
  Using Eq,~\ref{eq:pitch_angle_defn}, we computed the radial profiles of the pitch angle separately for stars and gas. Ideally, it would be worthwhile to investigate how pitch angles for stars and gas vary with galactocentric radius. However, due to the relatively coarse spatial resolution of TNG50-1 (compared to modern isolated $N$-body and $N$-body+SPH simulations), the pitch-angle profiles are noisy and hence pose a challenge to compare pitch-angle variations between stars and gas. Instead, we separately calculated the average pitch angle, $\tilde{i}_m$, for stars and gas, in a small radial bin of width $\Delta R = 0.5 \kpc$ around $R = 2.2 R_{\rm d}$. 
 The corresponding distribution of average pitch angle ($\tilde{i}_m$), for the $m=2$ and $m=4$ spirals in stars and gas, is shown in Fig.~\ref{fig:pitchangl_m2_m4_comparison_starsvsgas}. 
We also computed the Pearson correlation coefficient, $\rho$ to assess the correlation between stellar and gaseous spiral pitch angles. As shown in Fig.~\ref{fig:pitchangl_m2_m4_comparison_starsvsgas}, the average $m=2$ and $m=4$ pitch angles in gas and stars are only weakly correlated (especially, for the $m=4$ spirals). Furthermore, it is worth checking whether spirals in one component (stars or gas) are more tightly wound than those in the other component, at least statistically.  
We found that pitch angles for $m=2$ spirals in gas is (statistically) more tightly-wound than those for the stars. In detail, the distribution of $\tilde{i}_2$ for stars has a median value of $32.9 \degrees$ whereas the distribution of $\tilde{i}_2$ for gas admits a median value of $22.7 \degrees$. However, the pitch angle of $m=4$ spirals display a similar distribution for pitch angles, the median value of $\tilde{i}_4$ for stars is $27.4 \degrees$ while the median value of $\tilde{i}_4$ for gas is $27.5 \degrees$. 
This trend in average pitch angle between stars and gas likely reflects the fact that both components respond to the same underlying gravitational potential, resulting in spiral structures with comparable morphology. However, local phase shifts and differing dissipation properties may still cause subtle differences in the detailed structure of spirals in the two components. Moreover, since the gas is dynamically colder and can more easily dissipate energy, it may realign more quickly with spiral perturbations in the stellar potential, further reinforcing the morphological similarity.
A detailed investigation of radial profiles of spiral pitch angle in stars and gas will be taken up in a future study using isolated $N$-body+SPH simulations.

\subsection{Nature of spirals and associated chemical inhomogeneity}
\label{sec:age_chemistry_dissection}

\begin{figure*}[]
\includegraphics[width=1.012\linewidth]{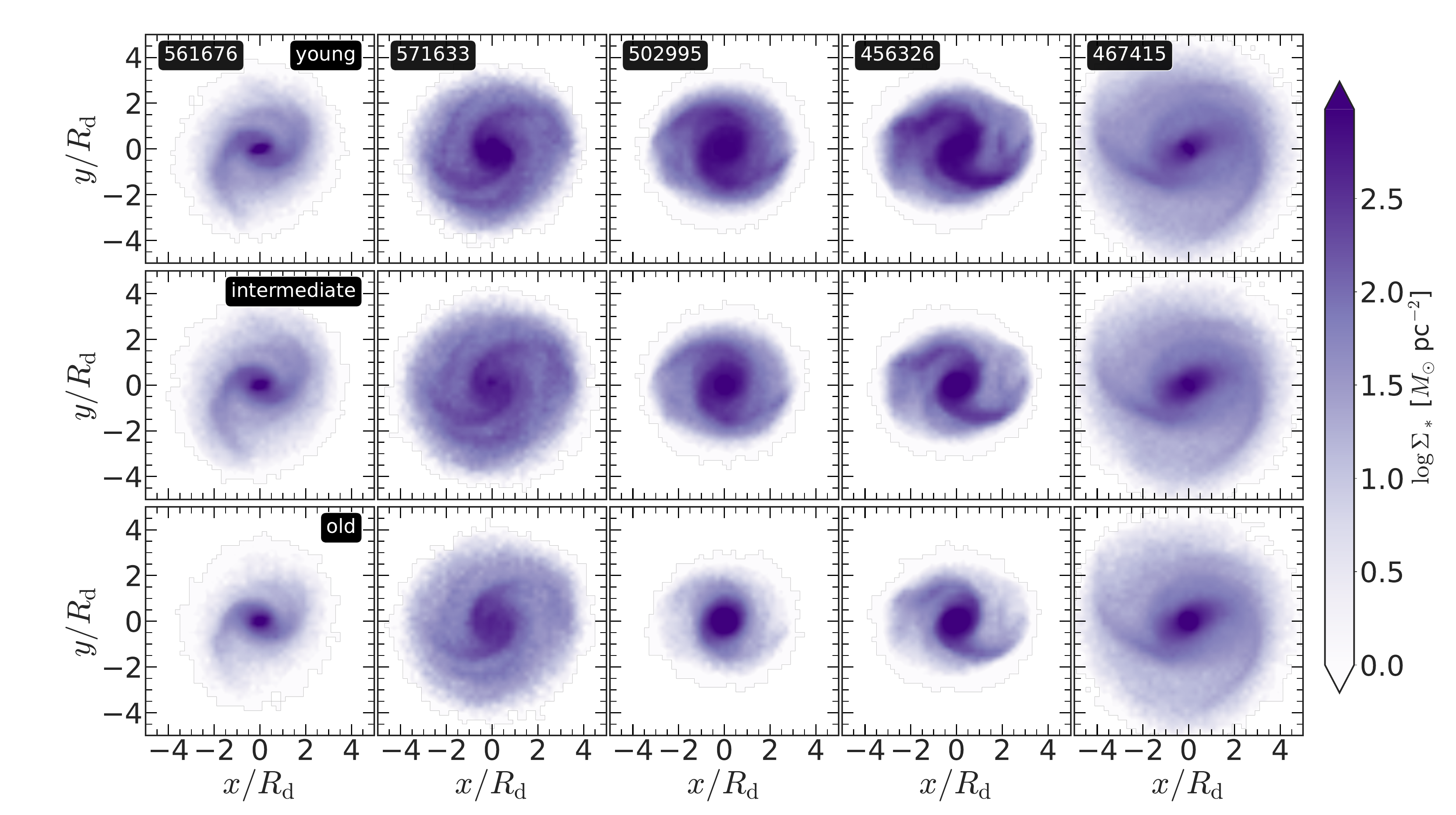}
\caption{\textit{Age dissection of spiral morphology:} Face-on ($x-y$-plane) density distribution of the stars of different age bins for a sample of non-barred (\textit{left two rows}) and spirals+bar (\textit{right two rows}) galaxies, considered for this work. \textit{Top panels} show for the young ($0 < \tau/\Gyr \leq 4$) stars, while \textit{middle} and \textit{bottom} panels show for the intermediate ($4 < \tau/\Gyr \leq 8$) and old ($8 < \tau/\Gyr \leq 12$) stars, respectively. $R_{\rm d}$ denotes the scale length of the stellar disk. The color bars show the corresponding surface density (in logarithmic scale). Spirals are present in all stellar populations of different age bins, indicating the density wave nature of spirals.}
\label{fig_densXYmap_ageDissection}
\end{figure*}

\begin{figure}
\includegraphics[width=0.95\linewidth]{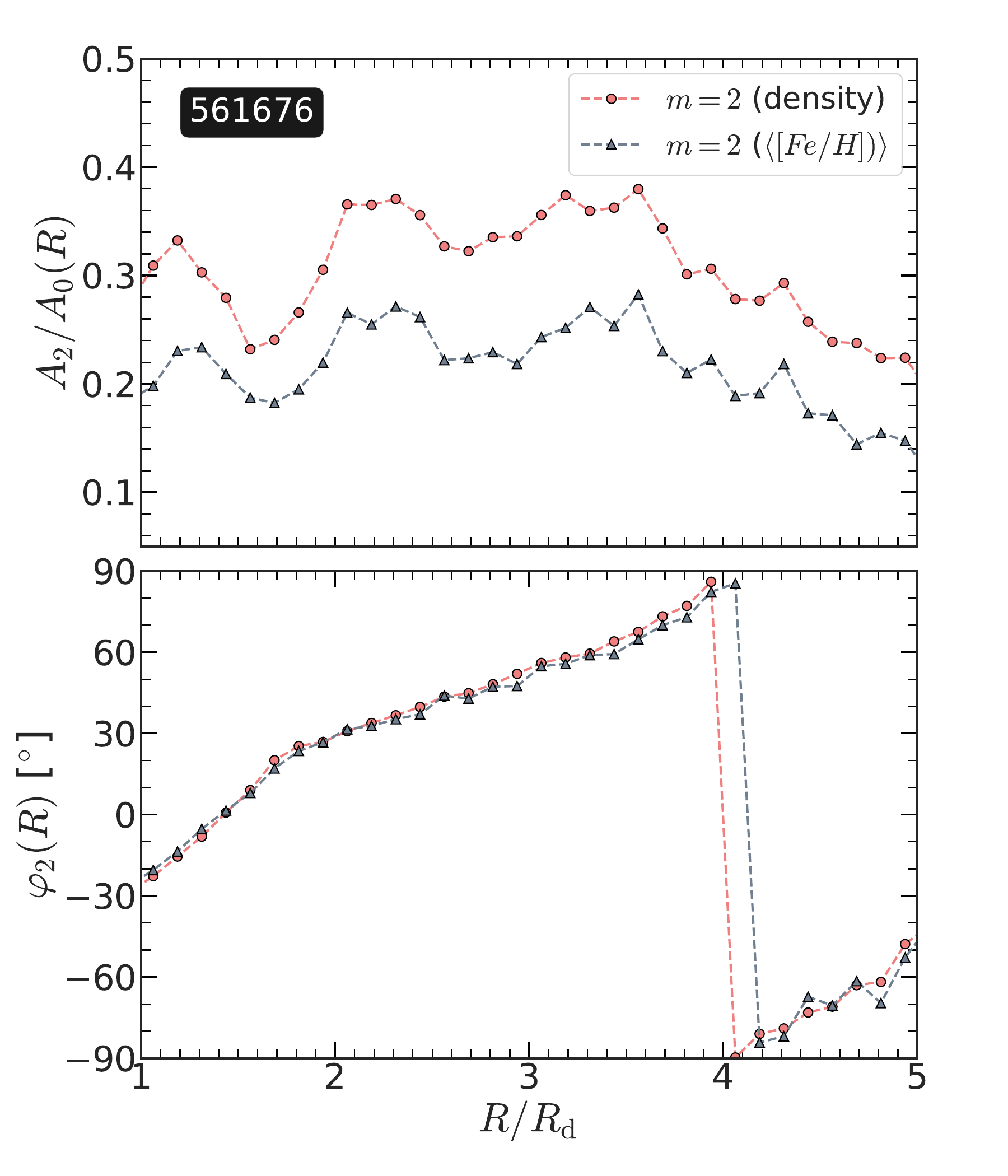}
\caption{Radial variation of the amplitude of the $m=2$ Fourier moment (\textit{top panel}), calculated from the density and the [Fe/H] distribution of stars (see the legend), and the corresponding phase angle, $\varphi_2$ (\textit{bottom panel}), for a non-barred MW/M31 analog. The spirals in chemistry of stars closely follow the spiral density wave in the stellar density distribution. }
\label{fig:spiralAmplitude_densVschem}
\end{figure}

In previous sections, we quantified some of the key properties of spirals in our selected sample of MW/M31 analogs. Here, we investigate further the nature of spirals and whether they leave imprints on the chemical composition of stars. To achieve that, we first calculated the age of stars, $\tau$, and then divided the stellar particles into three distinct age populations, namely, the young stars ($0 < \tau/\Gyr \leq 4$), intermediate stars ($4 < \tau/\Gyr \leq 8$) and old stars ($8 < \tau/\Gyr \leq 12$). Fig.~\ref{fig_densXYmap_ageDissection} shows the corresponding face-on density distribution of stars of different ages for a sample of non-barred and spirals+bar MW/M31 analogs. A visual inspection of Fig.~\ref{fig_densXYmap_ageDissection} reveals the presence of discernible spirals in all three stellar populations of different ages. The spirals are weaker in the old stellar population as compared to spirals in the young stellar population (compare top \& bottom rows of Fig.~\ref{fig_densXYmap_ageDissection}).  We further checked that this trend holds true for almost all MW/M31 analogs considered in this work. 
This suggests that spiral structure involves all stellar populations. This clearly shows that all stellar populations take part in the spiral structure, as opposed to \textit{material spirals} present in only the young stellar populations \citep[for a detailed discussion, see][]{BinneyTremaine2008,DobbsandBaba2014}. In other words, the spirals in the MW/M31 analogs are consistent with a density wave origin. A similar approach of determining spiral's density wave nature by dissecting stars into different age bins was adopted in \citet{Ghoshetal2022}.
{The presence of spiral features across all stellar age bins confirms that the arms are not transient star-formation features but genuine dynamical waves; thereby supporting the density-wave interpretation of spiral structure in cosmological context.
\par
Furthermore, a recent study by \citet{Poggioetal2022}, using \gaia\ DR3 showed that in the MW, distribution of stellar chemistry displays a characteristic azimuthal variation (irrespectively of chemistry and age cuts), likely to be excited by spirals in the MW. Furthermore, another recent study by \citet{Debattistaetal2025}, using a fiducial $N$-body+SPH model, showed that the metallicity variations, similar to what is seen in the MW, are present in young and old stars, and are coincident with the
spiral density waves. Here, using our selected MW/M31 analogs, we investigate whether spiral-driven metallicity variations are a generic feature. In Appendix~\ref{append_chemistryDissection}, we show the face-on distribution of the mean metallicity ($\avg{[Fe/H]}$) and mean age ($\avg{\tau}$) of stars for a sample of non-barred and spirals+bar sample of MW/M31 analogs (see Fig.~\ref{fig:densXYmap_ChemistryDissection}) used in this work. Even a mere visual inspection of Fig.~\ref{fig:densXYmap_ChemistryDissection} reveals the presence of discernible spirals in both the $\avg{[Fe/H]}$ and $\avg{\tau}$ distribution of stars, in agreement with the recent work by \citet{Debattistaetal2025}. Next, we investigate whether properties of the spirals in the $\avg{[Fe/H]}$ distribution are connected, in general,  to properties of spirals in the stellar density. To achieve, we Fourier decomposed the $\avg{[Fe/H]}$ distribution, and computed the amplitude of the $m$th Fourier moment as
\begin{equation}
A_m/A_0 (\avg{[Fe/H]})= \frac{\sum_j [Fe/H]_j e^{i m\phi_j}}{\sum_j   [Fe/H]_j}\,,
\label{eq:fourier_calc_chemistry}
\end{equation}
\noindent where $[Fe/H]_j$ denotes the metallicity of the $jth$ stellar particle. In Eq.~\ref{eq:fourier_calc_chemistry}, we excluded stars younger than $2 \Gyr$ to make sure that the rich star formation is not driving the variations in the $\avg{[Fe/H]}$ distribution. The corresponding radial variations of the $m=2$ Fourier amplitude and phase-angle ($\varphi_2$), for a non-barred MW/M31 analog is shown in Fig.~\ref{fig:spiralAmplitude_densVschem}. The $m=2$ Fourier amplitude and the phase-angle, computed for the [Fe/H] distribution display similar radial variation as compared with those computed from the stellar density distribution. In addition, the $m=2$ phase-angle of the spirals in the [Fe/H] do not show any phase lag when compared to the $m=2$ phase angle measured from the stellar density. A similar trend is also seen for the $m=4$ Fourier amplitude and phase-angle as well. We checked that for all MW/M31 analogs considered here, the spirals in [Fe/H] distribution traces the underlying spirals in the stellar density distribution. This is in agreement with the findings of \citet{Debattistaetal2025}. This further demonstrates a dynamically distinct scenario as compared to the azimuthal variations-likely caused by the resonant phenomenon where due to stars librating about a resonance might be expected to have [Fe/H] peaks azimuthally displaced relative to the perturbation \citep[e.g. see][]{Grandetal2016,Khoperskovetal2018}. Thus, the tight alignment between density, metallicity, and age spirals indicates that chemical and age variations can serve as dynamical tracers of spiral density waves—an observationally testable prediction for Gaia DR4 and the Roman Space Telescope.

\section{Underlying physical factors influencing spirals}
\label{sec_physicalImpacts}

\subsection{Gas content of galaxies}
\label{sec:gas_factor}

\begin{figure}
\includegraphics[width=0.95\linewidth]{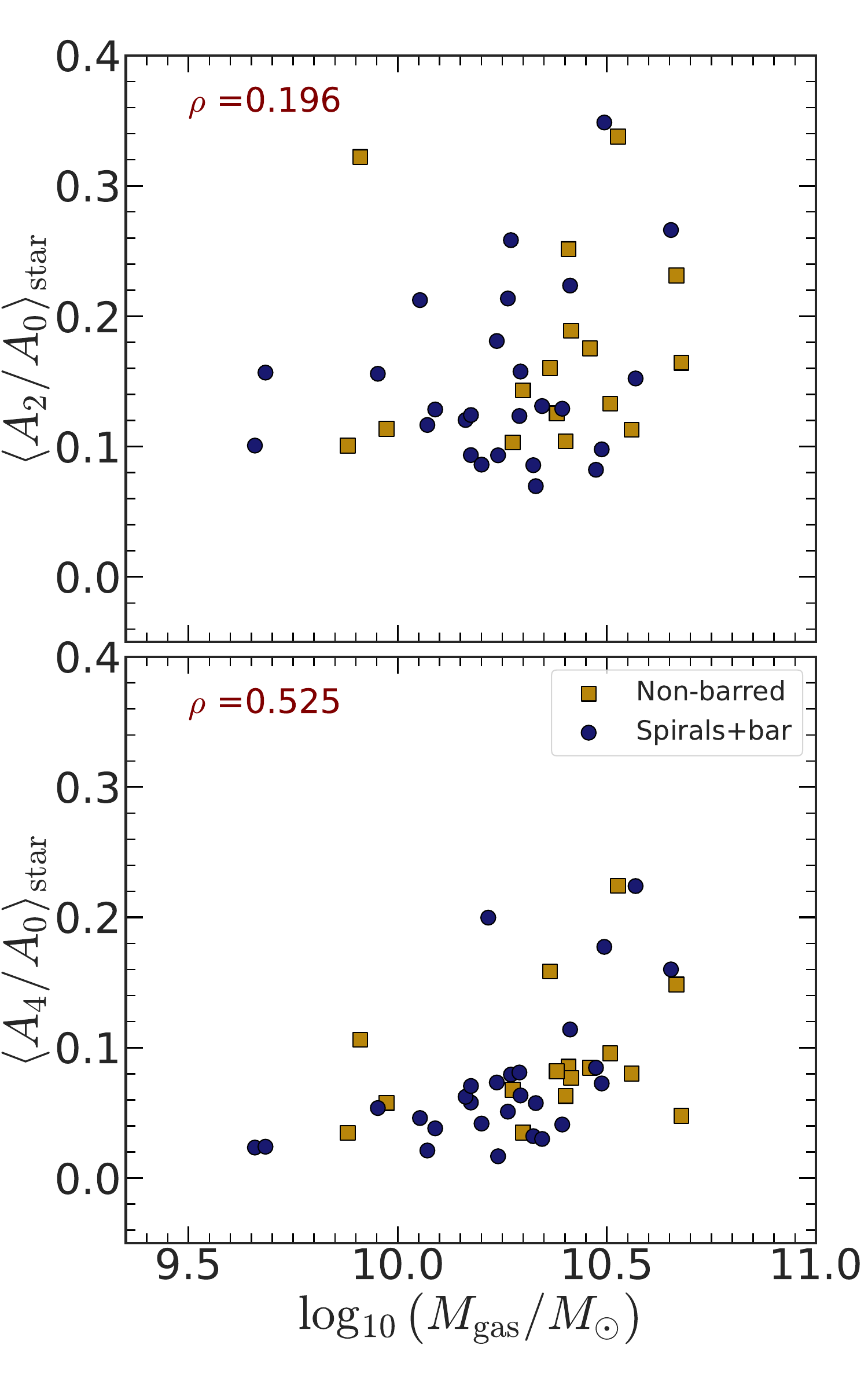}
\caption{Variation of the $m=2$ (\textit{top panel}) and the $m=4$ (\textit{bottom panel}) spiral strength as a function of gas mass, respectively. The blue filled circles denote the spirals+bar galaxies whereas the yellow filled squares denote the non-barred galaxies from our selected MW/M31 analogs. The Pearson correlation coefficient, $\rho$ is computed in each case, and shown at the top of each sub-panels.}
\label{fig_gasMass_vs_spiralAmplitude_m2m4}
\end{figure}

\begin{figure}
\includegraphics[width=0.95\linewidth]{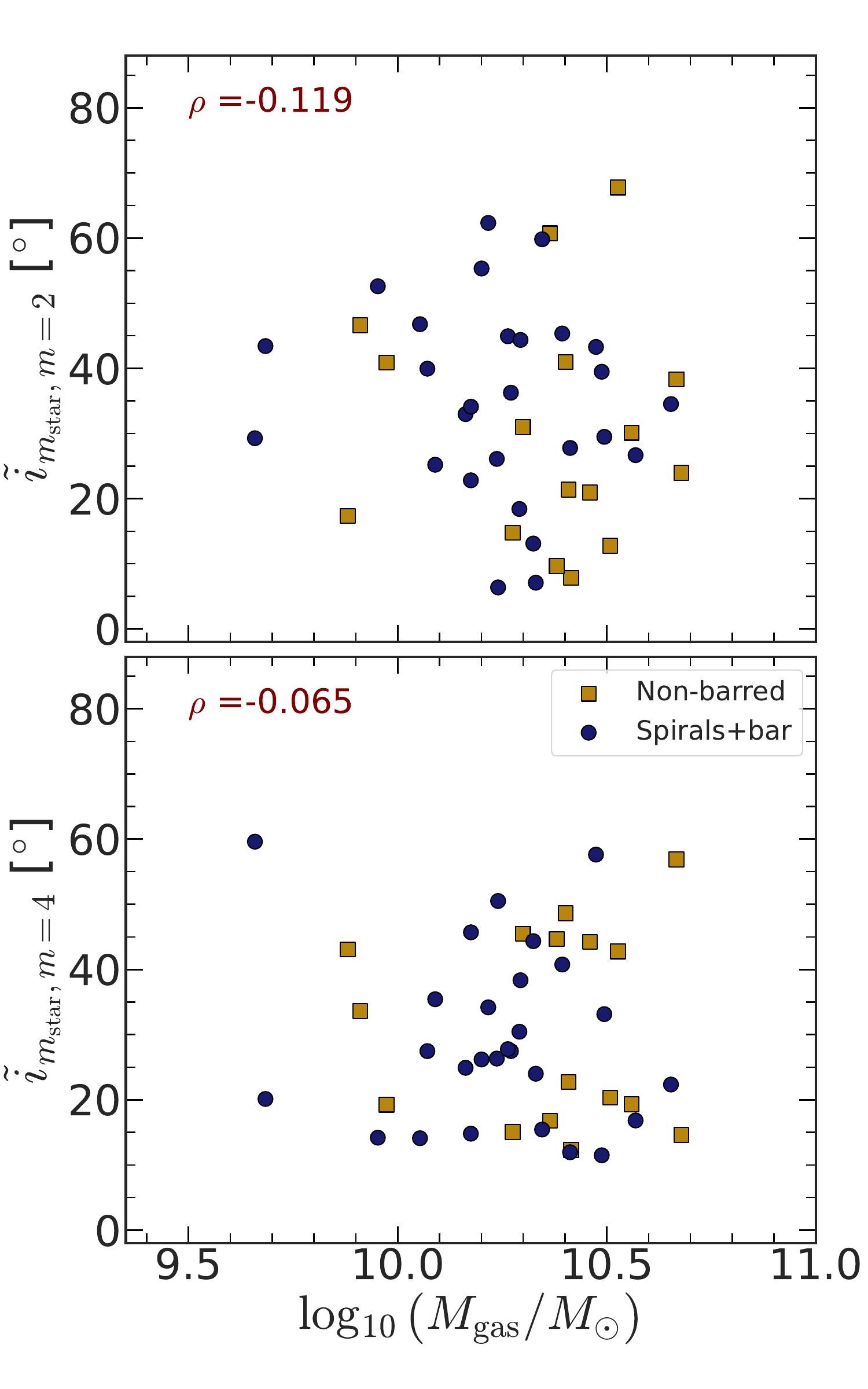}
\caption{Variation of the average pitch angle, $\tilde{i}_m$, for the  $m=2$ (\textit{left panel}) and the $m=4$ (\textit{right panel}) spirals as a function of gas mass, respectively. The blue filled circles denote the spirals+bar galaxies whereas the yellow filled squares denote the non-barred galaxies from our selected MW/M31 analogs. The Pearson correlation coefficient, $\rho$ is computed in each case, and shown at the top of each sub-panels.}
\label{fig_gasMass_vs_pitchangle_m2m4}
\end{figure}

Spiral galaxies contain a significant amount of interstellar gas, with typical gas fractions ranging from ~10–20\% in early-type spirals (Sa–Sb) to $>30 \%$ in late-type spirals (Sc–Sd) \citep[e.g.][]{YoungandScoville1991,BinneyandMerrifield1998}.
The dynamical role of gas has been studied in various contexts in galactic dynamics. 
Past semi-analytic studies showed that, owing to their low sound speed (as compared to the velocity dispersion of stars), presence of gas makes a system more susceptible to  both local axisymmetric \citep[e.g. see][]{JogandSolomon1984,Jog1996,Rafikov2001} and non-axisymmetric \citep{Jog1992} perturbations. Furthermore, gas plays a pivotal role in excitation, longevity of spirals in disk galaxies \citep[e.g. see][]{Jog1992,GhoshJog2016,GhoshandJog2018,GhoshandJog2022}.
\par
Here, we carry out a detailed study to find whether the interstellar gas has any influence on deciding spiral properties of the MW/M31 analogs selected for this work. To achieve that, first we calculated the total gas mass for a MW/M31 analog by summing the mass up of gas particles which fall within $5 R_{\rm d}$ and within $|z| \leq 5 \kpc$. Next, we investigate whether the strength of $m=2,4$ spirals in stars show any characteristic trend with the total gas mass. This is shown in Fig.~\ref{fig_gasMass_vs_spiralAmplitude_m2m4}. Furthermore, we computed the Pearson correlation coefficient, $\rho$, in each of these cases to quantify the correlation (if any). As seen from Fig.~\ref{fig_gasMass_vs_spiralAmplitude_m2m4}, the $m=2,4$ spiral amplitudes indeed show a positive trend with the total gas mass, that is, galaxies with higher total gas mass tend to show stronger $m=2,4$ spirals in stars, especially for the $m=4$ spirals. This trend is further confirmed from the Pearson correlation coefficient values. 
Since in realistic galaxy model, like the MW/M31 analogs studied here, stars and gas are gravitationally-coupled via the Poisson equation, therefore, the presence of more amount of a dynamically-cold component, namely gas, makes the total system (stars+gas) more prone to spiral instabilities than a purely stellar disk. 
\par
Lastly, we investigate whether there is any correlation between the average pitch angle of $m=2,4$ spirals in stars, $\tilde{i}_{m, star}$ with the total gas mass of MW/M31 analogs. This is shown in Fig.~\ref{fig_gasMass_vs_pitchangle_m2m4}. Fig.~\ref{fig_gasMass_vs_pitchangle_m2m4} suggests a very weak negative correlations between these two quantities. To confirm this, we calculated the Pearson correlation coefficients ($\rho$) for each of these cases. The resulting values of $\rho$ indeed shows that the average pitch angle of $m=2,4$ spirals in stars show almost no correlation or a very weak negative correlations for the MW/M31 analogs. 
Past semi-analytic work by \citet{Jog1992} claimed that the presence of more amount of gas increases the range of pitch angle over which swing amplification can happen \citep[also see][]{GhoshandJog2018}, thus resulting in broader stellar spirals arms. Here, we do not find this trend. However, we mention the fact that the values of $\tilde{i}_{m, star}$ only capture the essence of pitch angle in an average manner. A detailed study of radial variation of pitch angle of spirals in stars would be more insightful which we defer to future work due to the limited spatial resolution of TNG50-1 simulations. These results provide quantitative confirmation that a dynamically cold gas layer rejuvenates spiral instabilities, sustaining non-axisymmetric patterns over cosmic time.

\subsection{Galactic shear rate}
\label{sec:galactic_sheer}

\begin{figure}
\includegraphics[width=1.\linewidth]{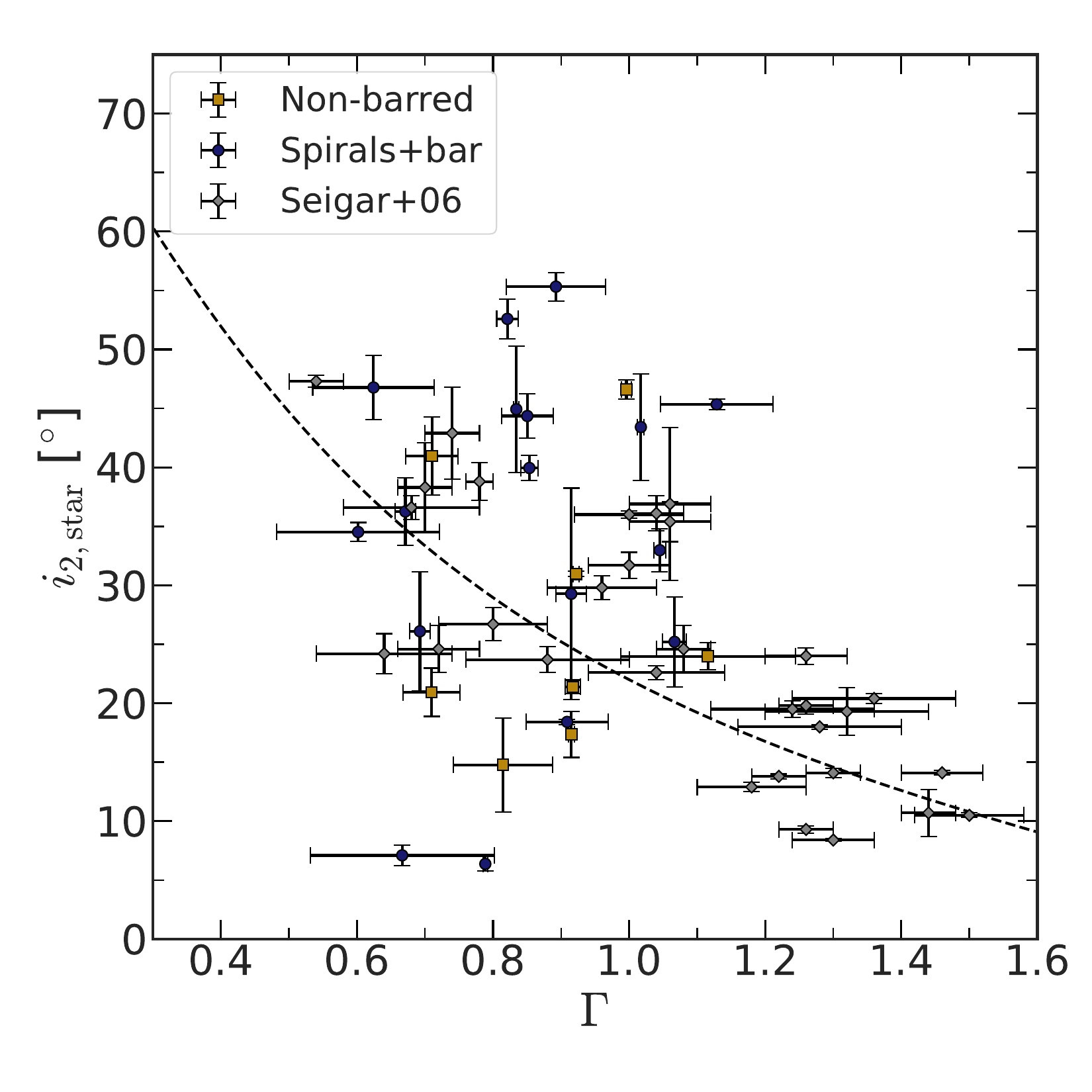}
\caption{Variation of the pitch angle of the $m=2$ spirals ($i_{2, \rm star}$) as a function of shear rate, $\Gamma$, both calculated within 1-kpc bin around $2.2 R_{\rm d}$, for the non-barred and spirals+bar MW/M31 analogs (see the legend) considered in this work. The black dashed line denotes the theoretical prediction from \citet{Michikoshi2014}. The error bars denote the standard deviation around the mean values. Grey points are taken from \citet{Seigaretal2006}.}
\label{fig:sshear_vs_pitchAngl}
\end{figure}

The differential rotation observed in typical spiral galaxies introduces a shear in the galactic disk \citep{BinneyTremaine2008}. Past theoretical studies have extensively examined the dynamical role of shear in the formation of swing-amplified spirals \citep[]{JulainToomre66,Toomre81,Jog1992,GhoshandJog2018}. In addition, the influence of galactic shear on spiral pitch angle has been extensively studied using both semi-analytic models and numerical simulations \citep[e.g. see][]{Grandetal2013,Michikoshi2014,Fujiietal2018} and from observations \citep[e.g.][]{Seigar2005,Seigaretal2006}. Here, we investigate the correlation (if any) between the spiral pitch angle and the galactic shear in the non-barred and spirals+bar MW/M31 analogs. At a given radius $R$, the galactic shear rate $\Gamma$ is defined as

\begin{equation}
\Gamma = - \frac{d \ln \Omega_{\rm circ}}{d \ln R}\,,
\label{eq:galactic_shear}
\end{equation}

 where $\Omega_{\rm circ}$ denotes the circular frequency of the disk at radius $R$.  The shear rate encapsulates the local differential rotation that governs how tightly a spiral winds; higher shear typically leads to more tightly wound arms.
 First, we calculated the net circular velocity, $v_{\rm circ}$, via the asymmetric drift correction \citep{BinneyTremaine2008} while using the stellar particles as kinematic tracers. The details are provided in Appendix~\ref{append_ADC}. 
 Next, using Eq.~\ref{eq:galactic_shear}, we calculated the radial variation of the shear rate, $\Gamma$, across the disk. For each galaxy, the calculation is performed from $R=1$ R$_{d}$ to $R=5$ $R_d$, ensuring coverage of both the inner and outer disk regions. The precise radial extent varies slightly between galaxies depending on the disk size and the quality of the circular velocity profiles. 
 The corresponding radial variation of $\Gamma$ for one such MW/M31 analog is also shown in Appendix~\ref{append_ADC} (see Fig.~\ref{fig:circvel_demo}). 
\par
We then examined whether the spiral pitch angle depends on the local shear rate. To achieve that, first we calculated the average values of the $m=2$ spiral pitch angle and $\Gamma$ in a radial bin of width $1 \kpc$ around $R=2.2 \ R_{\rm d}$. The specific choice of $2.2 \ R_{\rm d}$ is motivated by the choice of radial location made in past theoretical studies \citep[e.g. see][]{Seigaretal2006,Fujiietal2018}. 
This choice allows for a uniform comparison with previous studies. Fig.~\ref{fig:sshear_vs_pitchAngl} shows the relation between average shear rate and the $m=2$ spiral  pitch angle ($i_2$) for all the MW/M31 analogs considered in this work. For less than 5\% of our galaxy sample, the radial profiles of $v_{\rm circ}$ are noisy due to fluctuations in the $\sigma_R^2$ profile.
This, in turn, introduced large uncertainty/error in average $\Gamma$ value. These galaxies were excluded from the analysis shown in Fig.~\ref{fig:sshear_vs_pitchAngl}.
Furthermore, we compared our results with the theoretical prediction proposed by \citet{Michikoshi2014} having the function form 

\begin{equation}
 \tan i = \frac{2}{7} \frac{ \sqrt{4-2 \Gamma} }{\Gamma}\,.
 \label{eq:shear_scaling_relation}
\end{equation}

As seen in Fig.~\ref{fig:sshear_vs_pitchAngl}, the MW/M31 analogs exhibit noticeable scatter around the theoretical relation between the pitch angle and shear rate (see Eq.~\ref{eq:shear_scaling_relation}). 
This suggests that additional dynamical factors—such as bar influence, gas dynamics, or feedback—may modulate the relationship between pitch angle and shear rate beyond what is predicted by linear theory.
The scatter around the analytic pitch–shear relation (Eq.~\ref{eq:shear_scaling_relation}) is about 0.25 dex, reflecting modulation by bars, gas inflows, and feedback processes that locally alter angular-momentum transport beyond linear expectations.
In particular, bars can inject angular momentum and perturb the local disk dynamics \citep{Beaneetal23}, while gas dissipation and external accretion may influence the effective shear felt by the spiral structure. These complexities are expected in fully cosmological disks, and future high-resolution simulations may help isolate their individual contributions.

\subsection{Dynamical coldness of the disk}
\label{sec:stellarDMmass}

\begin{figure}
\includegraphics[width=\linewidth]{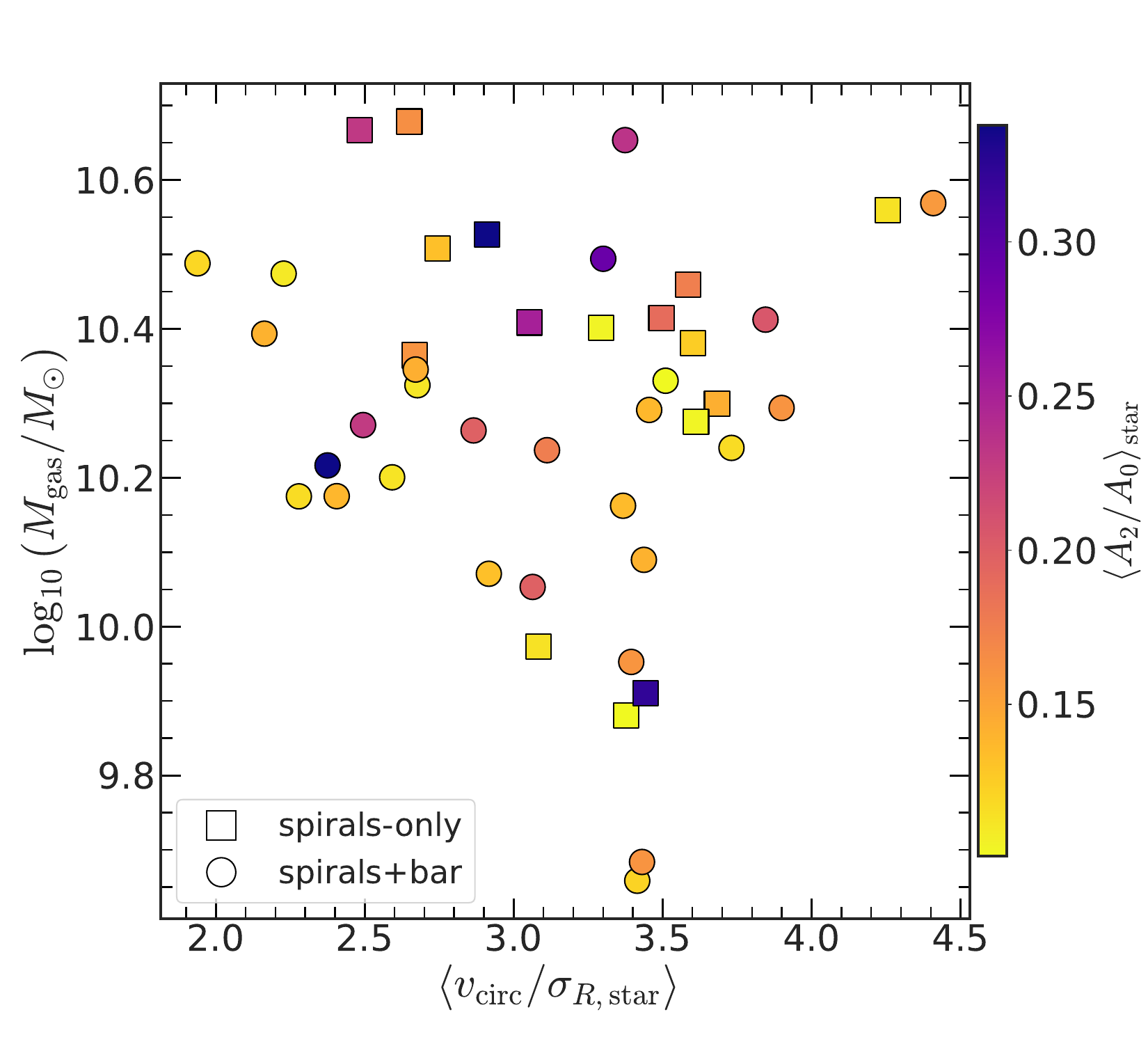}
\caption{Distribution of MW/M31 analogs as a function of total gas content (in logarithmic scale) and the average $v/\sigma$ value, \avg{v_{\rm circ}/\sigma_R}, color coded by the average strength of the $m=2$ spirals. Here, $v_{\rm circ}$ is net circular speed (from Eq.~\ref{eq:asy_drift1}), and $\sigma_R$ is the radial velocity dispersion for stars.  The values of \avg{v_{\rm circ}/\sigma_R} are calculated from the radial profiles of $v_{\rm circ}/\sigma_R$, and averaging is done within $2-4 \ R_{\rm d}$.}
\label{fig:voversigma_vs_spiralArm}
\end{figure}

Growth of any gravitational instability (e.g., spirals, bar) depends on the dynamical coldness of the galactic disk as well. Toomre Q parameter is often used to quantify the stability of a galactic disk against \textit{local, axisymmetric} perturbations \citep[e.g. see discussion in][]{BinneyTremaine2008}. Furthermore, the quantity $v/\sigma$ where $v$ denoting the ordered motion and $\sigma$ denoting the velocity dispersion respectively, is routinely used to understand the relative contribution of the internal systematic rotation and the random motion in balancing the dynamical equilibrium \citep[e.g. see][]{Illingworth1977,Binney1978,Daviesetal1983}. A lower value of $v/\sigma$ signifies that the galactic disk is more random motion dominated (or equivalently, dynamically hot) while a higher value of $v/\sigma$ denotes the ordered motion is more prevalent than the random motion (or equivalently, dynamically cold).
\par 
Here, we systematically investigate whether dynamical coldness of the galactic disk and the total gas content (being a lower velocity dispersion component) has any influence in deciding the strength of the spirals (or equivalently, the growth of the spiral instability). First, we computed the radial profiles of $v_{\rm circ}/\sigma_R$ where $v_{\rm circ}$ is the circular velocity (using Eq.~\ref{eq:asy_drift1}) and $\sigma_R$ is the stellar radial velocity dispersion. The average value of $v_{\rm circ}/\sigma_R$ is then calculated by taking the mean of $v_{\rm circ}/\sigma_R$ in the radial range $2-4 \ R_{\rm d}$. The resulting distribution of average strength of $m=2$ spirals ($\avg{A_2/A_0}$), as a function of $\avg{v_{\rm circ}/\sigma_R}$ and total gas mass is shown in Fig.~\ref{fig:voversigma_vs_spiralArm}. As seen from Fig.~\ref{fig:voversigma_vs_spiralArm}, the prominent spirals appear preferentially when the gas content or $\avg{v_{\rm circ}/\sigma_R}$ is higher or both the quantities have moderately intermediate values. In particular, no prominent $m=2$ spirals occur when $\langle v_{\rm circ}/\sigma_R \rangle < 3$ and $\log_{10}(M_{\rm gas}/M_\odot) < 9.9$, establishing quantitative thresholds for spiral growth in MW-mass galaxies. 
This further outlines the important roles of disk coldness and the gas content in the context of growth of spiral instabilities in MW-like galaxies.

\section{Discussion}
\label{sec:discussion}

The TNG50 Milky Way– and M31–mass analogs reveal that spiral structure persists even within galaxies subject to continuous accretion, feedback, and mergers. This demonstrates that long-lived density waves can coexist with cosmological growth, maintaining ordered patterns despite stochastic inflows. The combination of bars, gas inflows, and disk self-gravity sustains recurrent, quasi-stationary spiral modes over gigayear timescales, offering a cosmological confirmation of the mechanisms traditionally explored in isolated models (e.g., \citealt{Toomre81, SellwoodCarlberg1984,Dobbsetal2010, Donghia2013, Grandetal2013, Fujiietal2018}). These results place classical ideas of swing amplification and bar-driven spirals in a fully self-consistent cosmological setting.
\par 
The first main result is that most MW/M31 analogs harbor dominant two-armed ($m=2$) spirals in both the stellar and gaseous components, while the gaseous disks exhibit a broader distribution of multiplicities extending to $m>4$. The systematically higher amplitudes and tighter winding of gaseous spirals highlight the regulatory role of the cold interstellar medium, consistent with semi-analytic and numerical predictions that a dynamically cold gas layer rejuvenates stellar spiral instabilities and amplifies swing responses \citep{Jog1992, Rafikov2001, GhoshandJog2018, GhoshandJog2022}. Here, for the first time, this behavior is confirmed statistically for a cosmological sample of MW-mass galaxies.
\par 
A second robust finding is that spiral features appear across stellar populations of all ages, confirming that the arms are genuine dynamical waves rather than transient star-formation patterns. The alignment between spiral structure in stellar density, metallicity, and mean stellar age (see Fig.~\ref{fig:spiralAmplitude_densVschem}) demonstrates that chemical and age spirals trace the same phase as the density waves. This agrees with both $N$-body+SPH models and observations of the Milky Way \citep{Filionetal23,Debattistaetal2025}, where azimuthal variations in \feh\  are coincident with spiral arms. Such correlations provide an observationally testable prediction for \gaia~DR4 and the \textit{Roman}~Wide Field Imager.
\par 
A third key trend concerns the dependence of spiral amplitude on gas content and the dynamical temperature of the disk. Galaxies with higher gas fractions and larger values of $\langle v_{\rm circ}/\sigma_R\rangle$ display stronger $m=2$ and $m=4$ Fourier components, consistent with expectations that dynamically cold, gas-rich disks are more prone to non-axisymmetric instabilities \citep{JogandSolomon1984,Jog1992,GhoshandJog2018,GhoshandJog2022}.
The correlation between spiral amplitude, gas fraction, and disk coldness suggests a self-regulating cycle: gas accretion and cooling promote spiral growth, while stellar feedback and heating temporarily damp it. This mechanism provides a natural dynamical pathway linking cosmological inflow with internal disk evolution.
\par 
We also find substantial scatter in the relation between spiral pitch angle and galactic shear compared to analytic and empirical expectations \citep{Seigar2005, Seigaretal2006, Michikoshi2014,Fujiietal2018}. In idealized disks, stronger shear produces tighter winding, but in the cosmological TNG50 sample this correlation weakens. The $\sim0.25$-dex dispersion around the theoretical relation reflects the combined influence of bars, gas inflows, and feedback that locally modify angular-momentum transport and the effective shear. Bars in particular redistribute angular momentum through resonances (\citealt{LyndenBellKalnajs1972, Athanassoula2012, donghiaandAguerri2020, Beaneetal23}), producing departures from the classical shear–pitch relation and contributing to the observed diversity of spiral geometries.
\par 
Taken together, our results portray spiral structure in MW-mass galaxies as a self-organized, recurrent phenomenon sustained by the interplay between gas accretion, disk cooling, and angular-momentum redistribution. By quantifying these effects in a statistically significant cosmological sample, this study bridges classical disk-stability theory with galaxy-formation models. It provides the first empirical link between gas fraction, kinematic temperature, and spiral amplitude under realistic cosmological conditions, offering a predictive framework for interpreting the diversity of spiral morphologies observed from the local Universe to $z\!\sim\!2$.

\section{Summary}
\label{sec:conclusion}

We carried out a systematic study of quantifying the diversity of spiral structures in a statistically meaningful sample of Milky Way– and M31–like galaxies, where the galactic disks evolve naturally within a fully cosmological framework. For this work, we chose a sample 16 non-barred and 27 spirals+bar MW/M31 analogs, harboring prominent spirals, from the TNG50-1 cosmological simulation. We investigated how the spiral properties vary within the selected sample of MW / M31 analogues. Furthermore, we investigated different underlying physical factors which might influence the properties of the spirals. Our main findings are listed below.

\begin{itemize}

\item{Most MW/M31 analogs host dominant two-armed ($m=2$) spiral patterns in their stellar disks irrespective of the presence of a bar with occasional higher-order ($m>2$) structures. A larger fraction of galaxies exhibit multi-armed spirals in the gas, consistent with the greater dynamical responsiveness of the cold interstellar medium}.

\item{The gaseous $m=2$ and $m=4$ spirals are statistically stronger and more tightly wound than their stellar counterparts. This reflects the lower velocity dispersion of the gas and its ability to amplify and sustain density waves through dissipation and coupling to the stellar potential.}

\item{Spiral features are present across stellar populations of all ages, confirming that the arms are genuine dynamical waves rather than transient star-formation patterns. Spiral density waves also imprint coherent variations in mean stellar metallicity and age that follow the same phase as the density structure.}

\item{ Galaxies with higher gas fractions and dynamically colder stellar disks show stronger $m=2$ and $m=4$ spiral amplitudes, whereas gas-poor or kinematically hot systems exhibit only weak features. The relation between spiral pitch angle and galactic shear displays substantial scatter ($\sim 0.24$ dex) around the theoretically predicted relation, highlighting the combined influence of bars, gas inflows, and feedback on spiral morphology.}

\end{itemize}

Taken together, these findings establish the first statistical connection between gas content, disk kinematics, and spiral morphology in a cosmological context. They demonstrate that spiral density waves can persist and self-organize within galaxies undergoing continuous accretion and feedback, providing testable predictions for Gaia, JWST, and future Roman observations.

\begin{acknowledgments}
S.G. acknowledges funding from the IIT-Indore, through a Young Faculty Research Seed Grant (project: `INSIGHT'; IITI/YFRSG/2024-25/Phase-VII/02). The authors are grateful to the IllustrisTNG team for making their simulations publicly available. S.G. further thanks Eric Rohr and Annalisa Pillepich for their initial help in data handling of the TNG50 simulations.
\end{acknowledgments}

\begin{contribution}

Both the authors contributed equally in this manuscript.

\end{contribution}

%

\software{astropy \citep{2013A&A...558A..33A,2018AJ....156..123A,2022ApJ...935..167A},  
          }


\appendix

\section{Spirals in chemical and age distribution}
\label{append_chemistryDissection}

\begin{figure*}
\includegraphics[width=\linewidth]{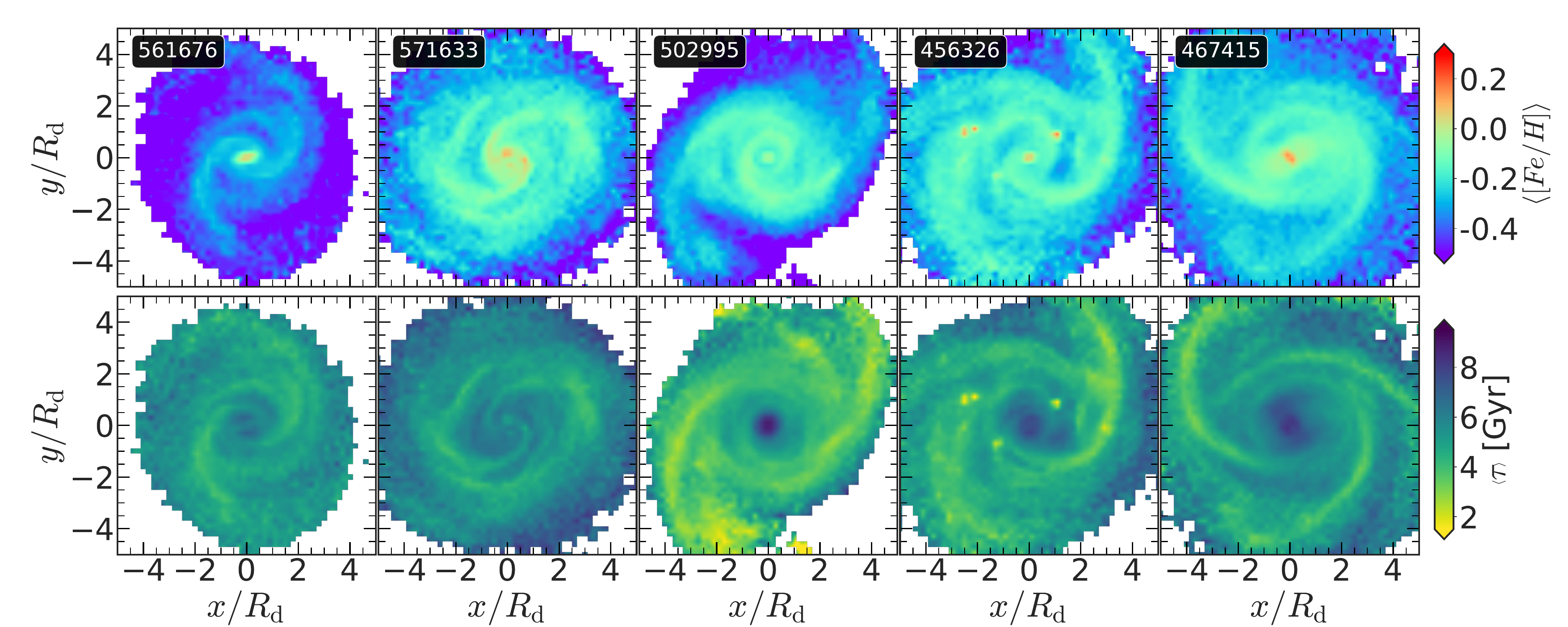}
\caption{\textit{Spirals in chemistry and mean age distribution:} Face-on ($x-y$-plane) distribution of the mean chemistry (\textit{top panels}) of stars ($\avg{[Fe/H]}$) and mean age (\textit{bottom panels}) of stars ($\avg{\tau}$) for a sample of non-barred (\textit{left three rows}) and spirals+bar (\textit{right two rows}) MW/M31 analogs. $R_{\rm d}$ denotes the scale length of the stellar disk. Spirals are present in the mean chemistry as well as in the mean age distribution of stars.}
\label{fig:densXYmap_ChemistryDissection}
\end{figure*}

Here, we calculated the distribution of average chemistry, $\avg{[Fe/H]}$, and mean stellar age, $\avg{\tau}$ of stars in a sample of non-barred and spirals+bar MW/M31 analogs used in this work. This is shown in Fig.~\ref{fig:densXYmap_ChemistryDissection}. To avoid the possibility that rich star formation driving the variations in the $\avg{[Fe/H]}$ distribution, we discarded stars of ages less than $2 \Gyr$ while calculating the face-on distribution of $\avg{[Fe/H]}$. For all MW/M31 analogs considered in Fig.~\ref{fig:densXYmap_ChemistryDissection}, both the distributions of  $\avg{[Fe/H]}$ and $\avg{\tau}$ show a prominent spiral feature.

\section{Calculation of circular velocity : Asymmetric drift correction}
\label{append_ADC}

\begin{figure}
\centering
\includegraphics[width=0.9\linewidth]{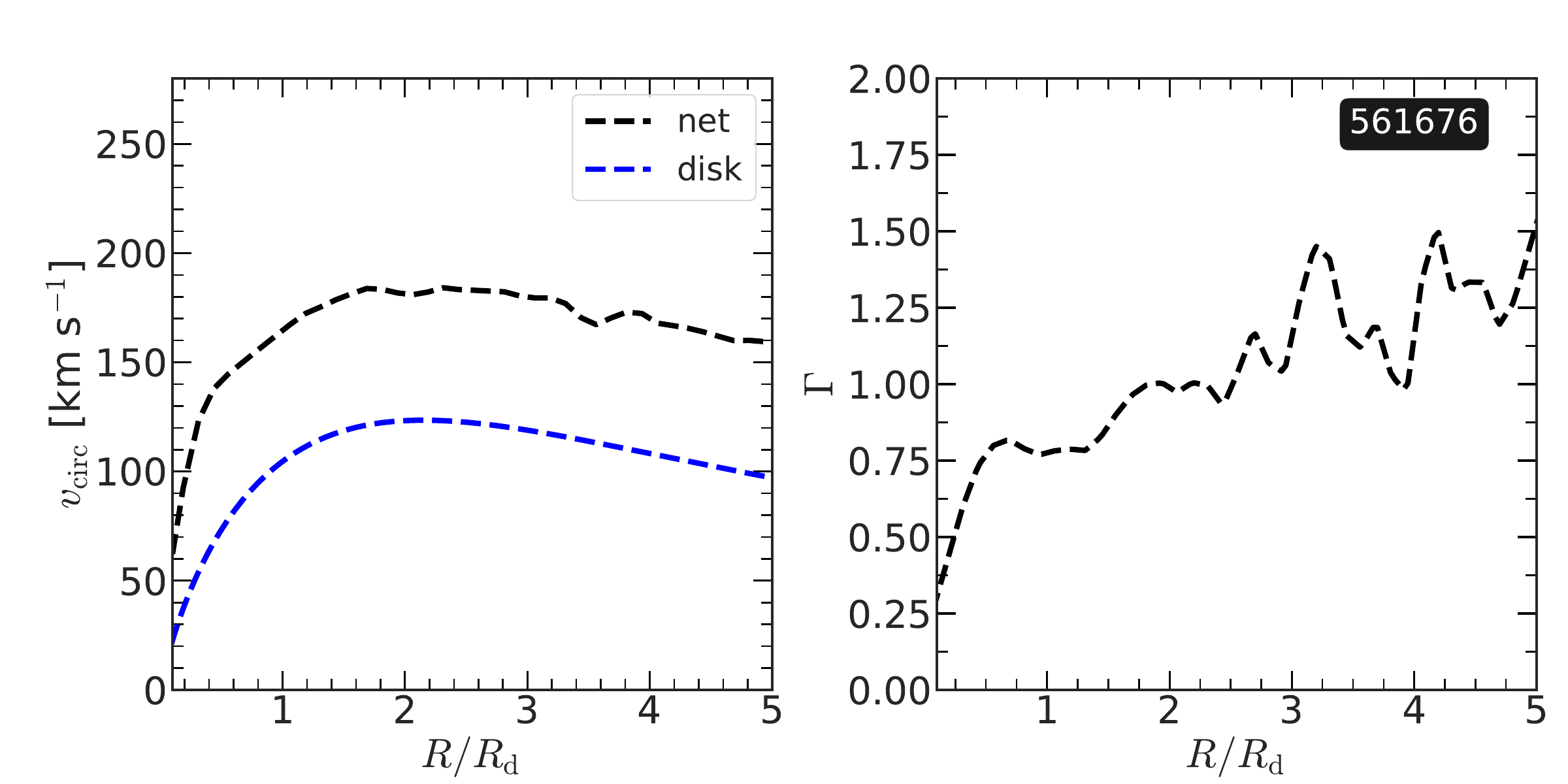}
\caption{\textit{Left panel:} Radial profile of the net circular velocity, $v_{\rm circ}$, calculated using the asymmetric drift correction (Eq.~\ref{eq:asy_drift1}), and the disk circular velocity, $v_{\rm c, disk}$ (Eq.~\ref{eq:vc_disk}) at $Z_r = 0$ for a non-barred MW/M31 analog. \textit{Right panel:} Radial profile of the galactic shear rate, $\Gamma$ (see Eq.~\ref{eq:galactic_shear}). $R_{\rm d}$ denotes the disk scale length. The $\texttt{SubHaloID}$ is mentioned at the top of the figure. }
\label{fig:circvel_demo}
\end{figure}
Here we briefly describe the calculation of circular velocity ($v_{\rm circ}$) by using the asymmetric drift correction \citep[for further details, see][]{BinneyTremaine2008} while using stars as the dynamical tracers. In particular, we used stars within $|z|/\kpc \leq 2$ for deriving the circular velocity in the mid-plane ($z=0$). First,  for a given galaxy in our selected sample, we computed the radial profiles of the stellar density ($\rho$), the azimuthal velocity ($v_{\phi}$) and the associated velocity dispersion components along the radial and the azimuthal directions ($\sigma_{R}$, $\sigma_{\phi}$), respectively. The circular velocity ($v_{\rm circ}$), at a given radius $R$, can be calculated from these quantities while correcting for the asymmetric drift via the equation 
\citep{BinneyTremaine2008}
\begin{equation}
v_{\rm circ}^2 = v_{\phi}^2+ \sigma_{\phi}^2 -\sigma_{R}^2 \left(1 + \frac{\rm d \ ln \ \rho}{\rm d \ ln \  R} + \frac{\rm d \ ln \ \sigma^2_R}{\rm d \ ln \  R}\right)\,.
\label{eq:asy_drift1}
\end{equation}
The corresponding radial profile of circular velocity ($v_{\rm circ}$), calculated at $Z_r = 0$ for a non-barred MW/M31 analog is shown in Fig.~\ref{fig:circvel_demo} (see left panel). In addition, we calculated the circular velocity due to the stellar exponential disk, $v_{\rm c, disk}$ using \citep{BinneyTremaine2008}

\begin{equation}
v_{\rm c, disk} = 4 \pi G \Sigma_0 R_{\rm d} y^2 \ \left[I_0(y)K_0(y) - I_1(y)K_1(y) \right]\,,
\label{eq:vc_disk}
\end{equation}
\noindent where $\Sigma_0$ is the central surface density, $y = R/2R_{\rm d}$, and $I_n$ and $K_n$ are the modified Bessel functions (or order $n$) of first and second kind, respectively. The corresponding radial variation of $v_{\rm c, disk}$ for the same non-barred MW/M31 analog is shown in Fig.~\ref{fig:circvel_demo} (see left panel). The right panel of Fig.~\ref{fig:circvel_demo} shows the radial variation of the galactic shear rate, $\Gamma$ (Eq.~\ref{eq:galactic_shear}) for the same $\texttt{SubHaloID}$.


\bibliography{my_ref}{}
\bibliographystyle{aasjournalv7}



\end{document}